\documentclass[aps,pra,reprint,superscriptaddress,floatfix,showpacs,showkeys]{revtex4-2}
\usepackage{xcolor}
\usepackage{graphicx}% Include figure files
\usepackage{dcolumn}% Align table columns on decimal point
\usepackage{siunitx}
\usepackage{float}
\usepackage{amsmath}
\usepackage{adjustbox}
\usepackage{bm}% bold math
\begin{document}
\title{Focused ion beam polishing based optimization of high-Q silica microdisk resonators}

\author{Lekshmi Eswaramoorthy}
%\email{X`lekshmi.eswaramoorthy@monash.edu}
 \affiliation{Laboratory of Optics of Quantum Materials, Department of Physics, Indian Institute of Technology Bombay, Powai, Mumbai -- 400076, India}
 \affiliation{Department of Materials Science and Engineering, Monash University, Clayton, Victoria 3800, Australia}
 \affiliation{IITB-Monash Research Academy, Indian Institute of Technology Bombay, Powai, Mumbai -- 400076, India}
\author{Parul Sharma}
\author{Brijesh Kumar}
\author{Abhay Anand V S}
\author{Anuj Kumar Singh}
\author{Kishor Kumar Mandal}

%\email{X`lekshmi.eswaramoorthy@monash.edu}
 \affiliation{Laboratory of Optics of Quantum Materials, Department of Physics, Indian Institute of Technology Bombay, Powai, Mumbai -- 400076, India}

\author{Sudha Mokkapati}
 \email{sudha.mokkapati@monash.edu}
 \affiliation{Department of Materials Science and Engineering, Monash University, Clayton, Victoria 3800, Australia}
 \affiliation{IITB-Monash Research Academy, Indian Institute of Technology Bombay, Powai, Mumbai -- 400076, India}
 
\author{Anshuman Kumar}
\email{anshuman.kumar@iitb.ac.in}
 \affiliation{Laboratory of Optics of Quantum Materials, Department of Physics, Indian Institute of Technology Bombay, Powai, Mumbai -- 400076, India}
\affiliation{Centre of Excellence in Quantum Information, Computation, Science and Technology, Indian Institute of Technology Bombay, Powai, Mumbai 400076, India}
\date{\today}

\begin{abstract}
Whispering gallery mode (WGM) microdisk resonators are promising optical devices that confine light efficiently and enable enhanced nonlinear optical effects. This work presents a novel approach to reduce sidewall roughness in SiO\textsubscript{2} microdisk resonators using focused ion beam (FIB) polishing. The microdisks, with varying diameter ranging from 5 to 20 $\mu$m are fabricated using a multi-step fabrication scheme. However, the etching process introduces significant sidewall roughness, which increases with decreasing microdisk radius, degrading the resonators' quality. To address this issue, a FIB system is employed to polish the sidewalls, using optimized process parameters to minimize Ga ion implantation. 
White light interferometry measurements reveal a significant reduction in surface roughness from 7 nm to 20 nm for a 5 $\mu$m diameter microdisk, leading to a substantial enhancement in the scattering quality factor (Qss) from $3\times 10^2$ to $2\times 10^6$. 
These findings demonstrate the effectiveness of FIB polishing in improving the quality of microdisk resonators and open up new possibilities for the fabrication of advanced photonic devices.

\end{abstract}

\maketitle

\section{Introduction}

Whispering gallery mode (WGM) resonators are optical devices that leverage the phenomenon of waveguiding by curved boundaries, akin to acoustic whispering galleries. These resonators can achieve high quality factors, enabling efficient light-matter interaction, which combined with their compact design allows for enhanced nonlinear optical effects\cite{Lin2017}, crucial for applications in nonlinear and quantum optics\cite{Strekalov2016} and sensing\cite{Foreman2015}. WGM resonators can have various geometries that significantly influence their optical properties and applications, such as spherical, cylindrical, and ring configurations. These resonators are primarily characterized by their axisymmetric shapes, which facilitate the confinement of light through total internal reflection. 

Microdisks offer several advantages in optical applications, particularly in terms of quality factors, integration capabilities, and lasing efficiency.  Microdisks can be easily integrated with fiber optics and other photonic components, allowing for robust coupling and versatile applications in photonic integrated circuits \cite{seung_june_choi__2003}. The ability to parallel integrate multiple microdisks with a single fiber taper enhances their utility in complex systems. Microdisks can achieve exceptionally high quality factors (Q), which is crucial for applications requiring low loss and high sensitivity \cite{paul_e__barclay__2006}. While microdisks can achieve high Q values, they often face limitations that affect their performance in practical applications.
\begin{figure}[htbp]
    \centering
    \includegraphics[width=0.8\linewidth]{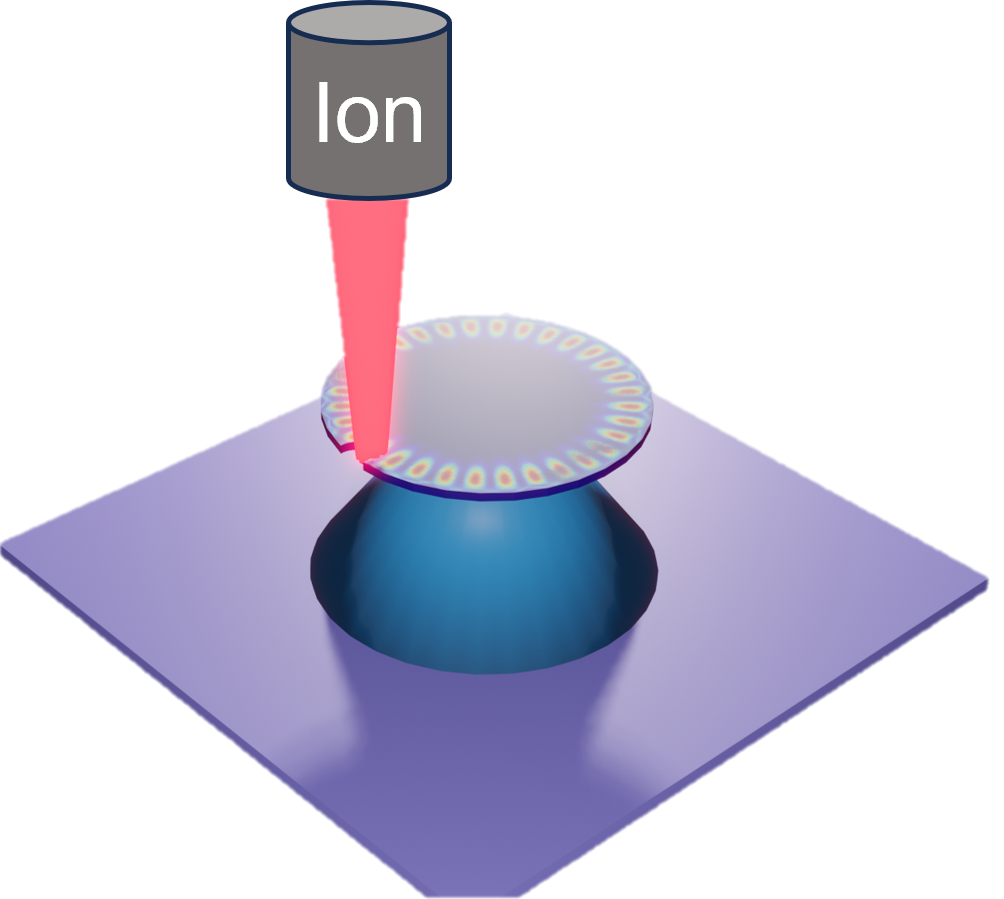}
    \caption{Schematic of sidewall polishing using Focused Ion Beam (FIB) technique}
    \label{fig:1}
\end{figure}

%Fabrication of microdisk resonators encompasses a variety of techniques and materials to achieve resonators with specific characteristics. Similarly, the fabrication of lithium niobate microdisk resonators with high optical quality factors using scalable techniques over a wide wavelength range has been showcased \cite{Wang:14}.

\begin{figure*}[ht]
    \centering
    \includegraphics[width=0.7\linewidth]{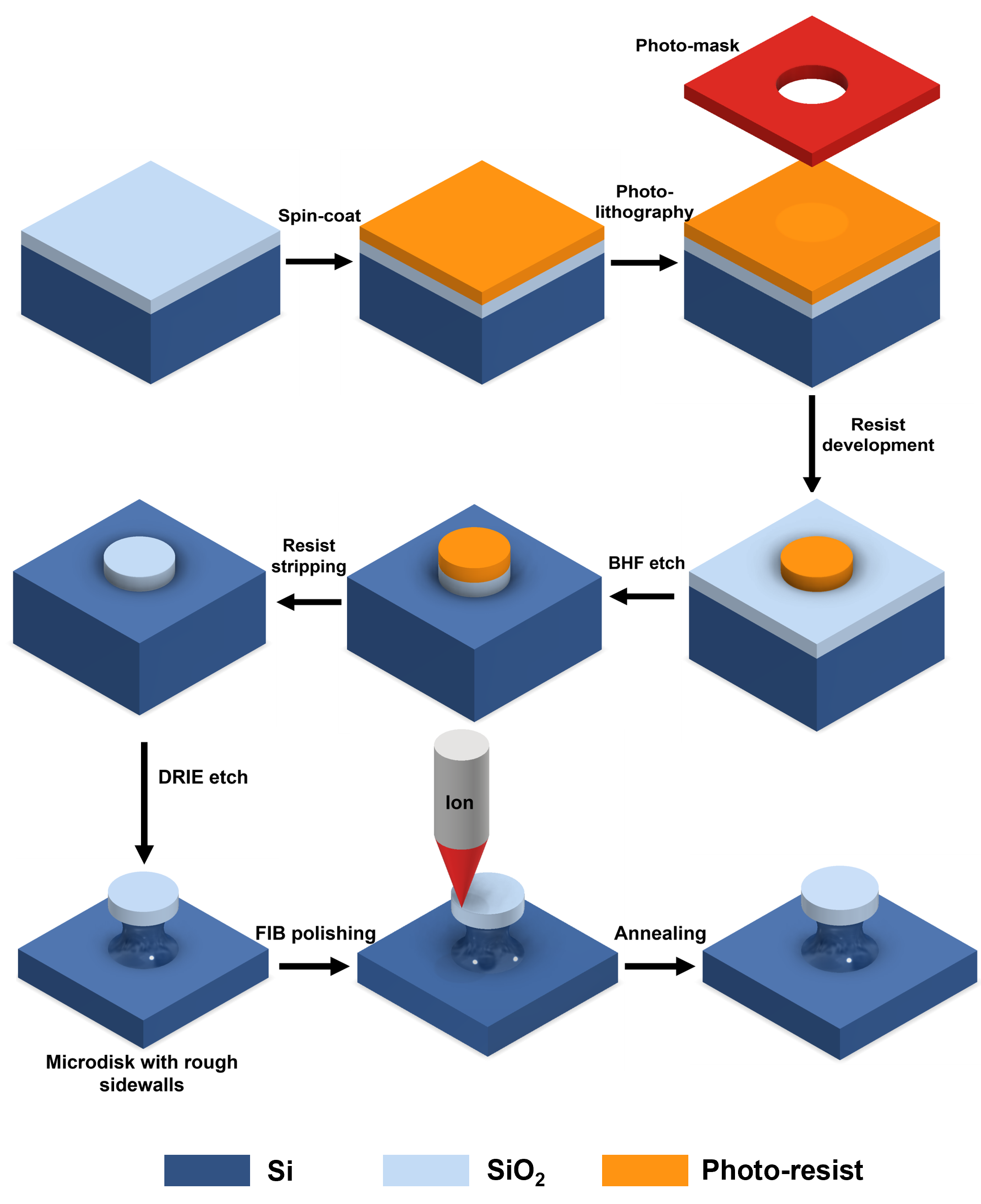}
    \caption{Fabrication process flow for Silica microdisk with smooth sidewalls}
    \label{fig:2}
\end{figure*}

Both subtractive as well as additive fabrication processes have been employed for microdisks \cite{9393ab6383934eff8776bb47f7287ee7}, on different material platforms. For instance, microdisk resonators were fabricated from an organic semiconductor material using vacuum vapor deposition and photolithography without dry etching \cite{doi:10.1021/acs.jpcc.1c02665}.  Electron-beam direct-writing was employed to fabricate microdisk resonators with high-quality factors \cite{https://doi.org/10.1002/anie.202007361}, ion slicing and wafer bonding for lithium niobate microdisk resonators \cite{Wang:14}, direct laser writing for polymeric microdisks on silicon \cite{Grossmann:11}, and wet etching for silica-based microcavities \cite{10.1063/5.0171764}. The fabrication of high-quality silicon optical microdisk resonators on silicon-on-insulator wafers was demonstrated using resist reflow and low-damage plasma dry etching \cite{Borselli:05}. These techniques offer various advantages in terms of scalability and ease of fabrication. Thus, the fabrication of microdisk resonators involves a range of materials and methods tailored to specific requirements such as high quality factors, tunability, and efficient coupling. While diverse approaches including resist reflow, dry etching, ion slicing, direct laser writing, and wet etching have been used to realize microdisk resonators with desired characteristics, most of these techniques result in sidewall roughness at multiple stages of the fabrication process. Interference fringes in photoresist used in lithography \cite{gabor_mezosi__2009}, formation of ripples during etching \cite{martial_chabloz__2000}, gas flow rate, platen power, etching cycle time \cite{h__c__liu__2003} can all be the parameters affecting sidewall roughness. 

\begin{figure*}[ht]
    \centering
    \includegraphics[width=1\linewidth]{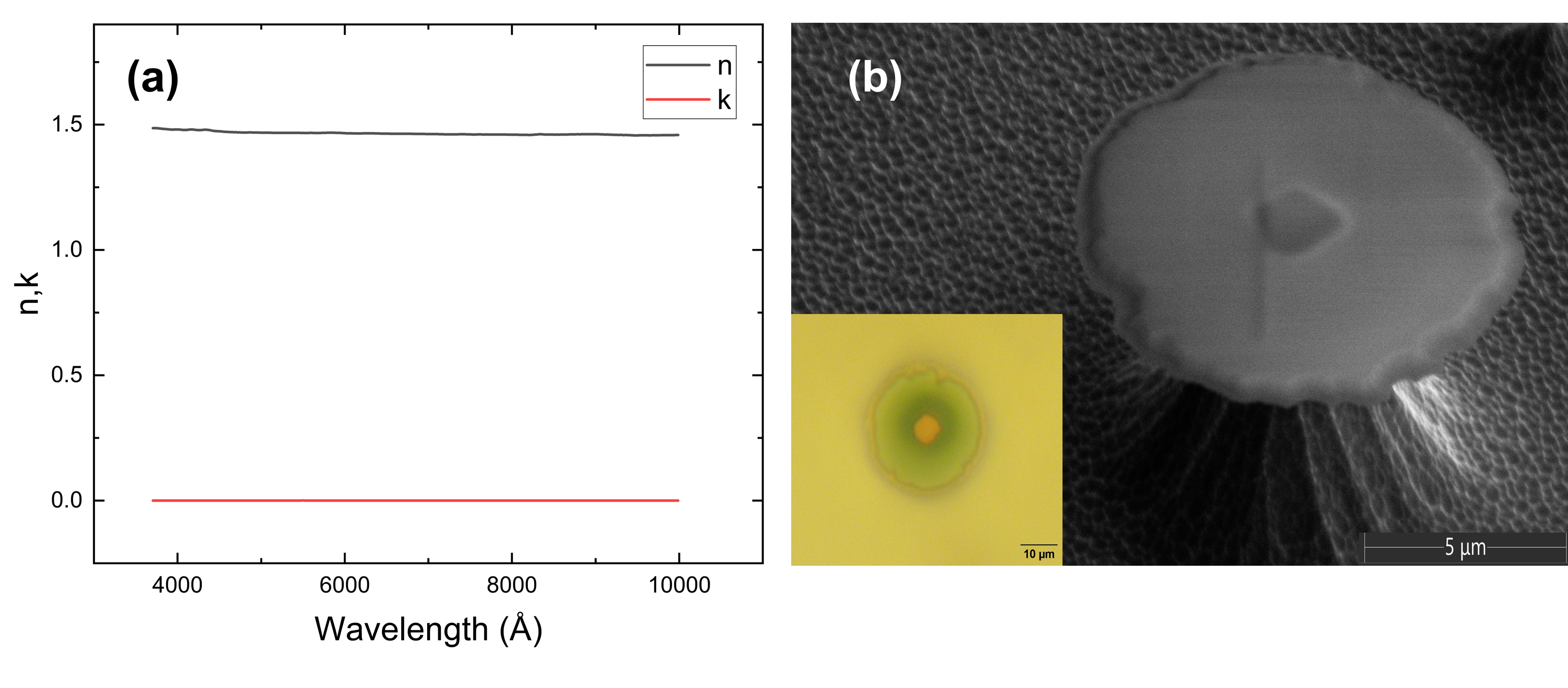}
    \caption{(a) Wavelength-dependent variation of the refractive index ($n$) and extinction coefficient ($k$) for the SiO\textsubscript{2} film deposited; (b) SEM image of the microdisk post DRIE, (inset) an optical micrograph of the same disk }
    \label{fig:3}
\end{figure*}

Sidewall roughness in microdisk cavities can be eliminated or reduced through various post-processing techniques and fabrication optimizations. The choice of method depends on factors such as the material used, the desired level of smoothness, and the specific requirements of the application. Chemo-mechanical polishing has been demonstrated to significantly reduce surface roughness in silica microdisks, resulting in ultra-high quality factors exceeding $10^8$ \cite{10.1063/5.0051674}. Thermal reflow techniques have shown promise in reducing sidewall roughness and associated optical scattering loss in high-index-contrast, sub-micron waveguides \cite{Hu:10}. Wet chemical polishing combined with dry etching has been successfully applied to GaN sidewalls, removing etching damages and smoothing vertical sidewalls \cite{He2018}. Femtosecond laser polishing has emerged as a flexible and non-contact method for post-processing additively manufactured parts, including sidewalls \cite{LI2023155833}. Resist reflow techniques, such as the chip-upending method, have been shown to reduce line edge roughness in submicron single-mode waveguides \cite{doi:10.1142/S0218863510005741}. However, these techniques generally require complex fabrication techniques and precise control of parameters that alter the morphology of the designed microcavity. In this work, we propose using a focused ion beam to deterministically reduce the roughness by polishing the sidewalls of the microcavity (Figure \ref{fig:1}). This method retains the shape and dimensions of the microcavity and allows for a sub-nanometer resolution of control of polishing.

\section{Methods}
The microdisk resonator is designed using FDTD and MODE tools (Ansys LUMERICAL) \cite{Eswaramoorthy_2022}. In this work, we select a resonator thickness $t$ of $300$ nm and a diameter $D$ ranging from $\SI{5}{\um}$ to $\SI{20}{\um}$ to support resonances in the visible wavelength range. The geometrical parameters of the microdisk are optimized to enable single-mode operation, tailoring the structure for potential future applications such as lasing.  The SiO$_2$ microdisk is fabricated as shown in Figure \ref{fig:2}. Thermally grown SiO$_2$ generally exhibits superior qualities compared to CVD-grown SiO$_2$, particularly for fabricating optical structures \cite{Gonzalez}. The thermal oxidation process produces a denser, more uniform, and higher-quality oxide layer with fewer defects and impurities \cite{PhysRevLett.93.036106}. Additionally, thermally grown SiO$_2$ typically contains fewer intermediate oxidation states, which can act as traps and affect optical performance \cite{Izumi1999}. In our case, the $SiO_2$ thin film is thermally grown on a 2-inch $Si$ substrate using a wet oxidation furnace. The  refractive index $n$ and extinction coefficient $k$ (shown in Figure \ref{fig:3} ) of the film were determined using an ellipsometer (J.A. Woollam M-2000). The wafer was then diced into $\SI{5}{mm} \times \SI{5}{mm}$ samples and spin-coated with a positive photo-resist (MICROPOSIT S1813) using a spin processor (Laurell WS-650-23NP). UV light exposure transfers the micro-disk pattern from the hard mask to the sample using a double-sided mask alignment system (EVG 620). The pattern is transferred onto the photo-resist film upon treating the sample with a developer (MICROPOSIT MF-321). The pattern is further transferred to the $SiO_2$ film by the wet etching technique using Buffered HF acid. The BHF has a more stable etch rate and is more gentle to photoresist due to an almost pH-neutral solution, which makes photoresist a good masking material for the oxide etch. \\

\begin{figure*}[ht]
    \centering
    \includegraphics[width=1\linewidth]{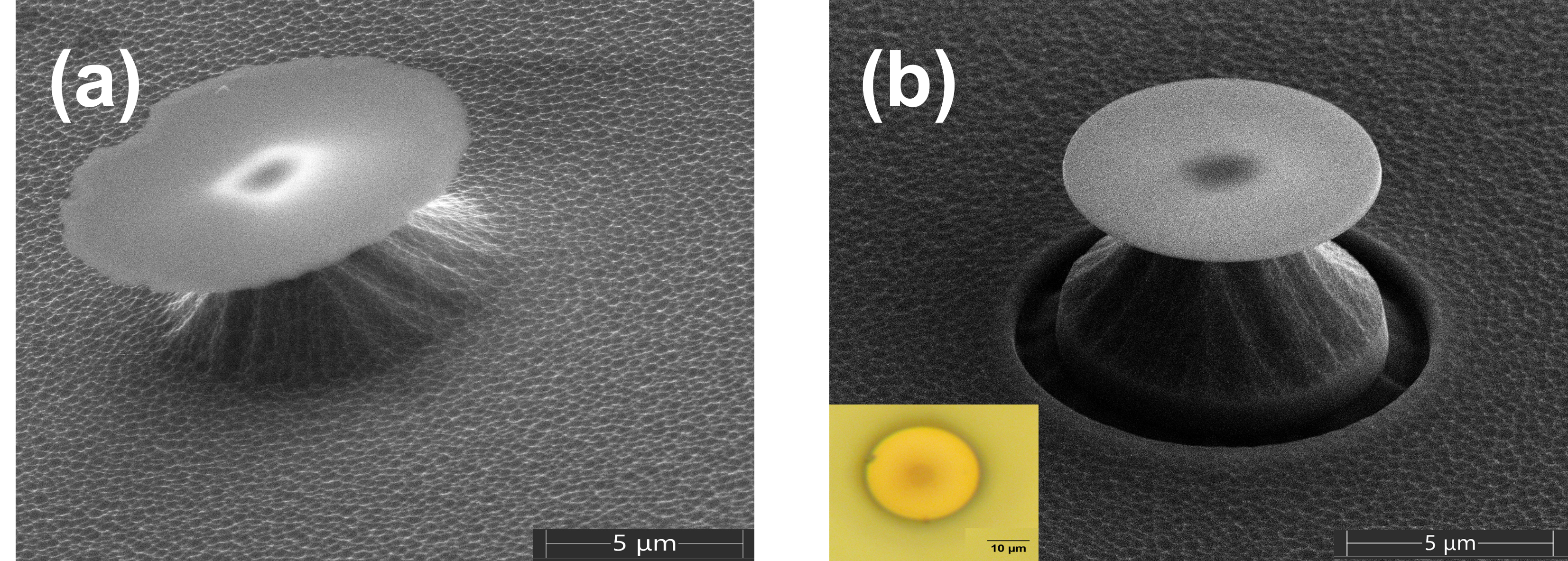}
    \caption{SEM images of Silica microdisk (a) post DRIE with rough sidewalls; (b) post FIB polishing with smooth sidewalls (inset) an optical micrograph of the polished disk }
    \label{fig:4}
\end{figure*}

The microdisk resonator is designed to be isolated to prevent any leakage loss into the $Si$ substrate. The isolation is obtained in the form of a pedestal by etching out the underlying $Si$. Wet etchants such as KOH and TMAH can be optimised to produce an isotropic etch profile, but the selectivity of these etchants towards $SiO_2$ over $Si$ is poor, and the resultant surface shows significant roughness \cite{Purohit_2022}. Thus, we use the dry etching route using the combination of $SF_6$ and $O_2$ gases in a Deep Reactive Ion Etching (DRIE) system (SAMCO RIE-400iPB). This inductively coupled plasma (ICP) RIE system uses high-density plasma to perform high-speed Silicon etching. Photoresist hardening during RIE at high ICP power is a complex phenomenon that can be attributed to several factors. The high ICP power leads to increased ion bombardment and plasma density, which can cause cross-linking of the photoresist polymer chains \cite{10.1116/6.0002109}. This cross-linking effect makes the photoresist more resistant to etching, effectively hardening it. The increased ion bombardment can also lead to the formation of a thin, hardened layer on the photoresist surface, known as the "skin effect" \cite{Ahn2004}. To prevent such issues, we remove the photo-resist before subjecting the sample to DRIE. The photo-resist already starts to delaminate from the edges upon its interaction with BHF and a further treatment using an N-methyl-2-pyrrolidone solvent (Kayaku Advanced Materials, Remover PG) completely strips off the photo-resist from the sample. The sample is now subjected to DRIE with the $SiO_2$ film as the hard mask. The optimised process parameters to obtain an isotropic etch profile to selectively etch $Si$ over $SiO_2$ is listed in Table \ref{tab:table_1}. 

\begin{table}[H]
    \centering
    \begin{adjustbox}{max width=\textwidth} % Adjusts table to fit within text width
    \begin{tabular}{|c|c|c|}
        \hline
        \textbf{Parameter} & \textbf{Step 1} & \textbf{Step 2} \\ \hline
        SF\textsubscript{6} Mass Flow (sccm) & 30 & 200 \\ \hline
        O\textsubscript{2} Mass Flow (sccm) & 300 & 0 \\ \hline
        Chamber Pressure (Pa) & 2 & 4 \\ \hline
        RF Bias Power (W) & 60 & 30 \\ \hline
        ICP Power (W) & 800 & 1000 \\ \hline
        Process Time (s) & 10 & 1 \\ \hline
    \end{tabular}
    \end{adjustbox}
    \caption{Process Parameters for Deep Reactive Ion Etching (DRIE) of $Si$ with $SiO_2$ hard mask}
    \label{tab:table_1}
\end{table}

For the given process parameters, a $\SI{5}{\um}$ high isolation was obtained for a microdisk of $\SI{5}{\um}$ radius. However, the microdisk was found to have significant sidewall roughness as evident from the SEM images shown in Figure \ref{fig:3}. We found that the roughness of the sidewalls increased as the radius of the microdisk decreased and this lead to the degradation in the quality of the resonators. We therefore incorporated an additional fabrication step, polishing the sidewalls to reduce roughness. 

\subsection{Focused Ion Beam Polishing}

Focused ion beam (FIB) systems have become a versatile tool for nano-machining and local modification of materials \cite{Gierak_2009}, \cite{Langford2007}. In particular, electronic \cite{gamo1996nanofabrication} and magnetic \cite{khizroev2003focused} structures can be fabricated directly by patterning the samples using a scanned beam. For these nanofabrication purposes, the feature resolution is determined by a variety of parameters, including the beam probe size, the beam profile and the distribution of primary ions in the target material. 

\begin{figure*}[ht]
    \centering
    \includegraphics[width=0.9\linewidth]{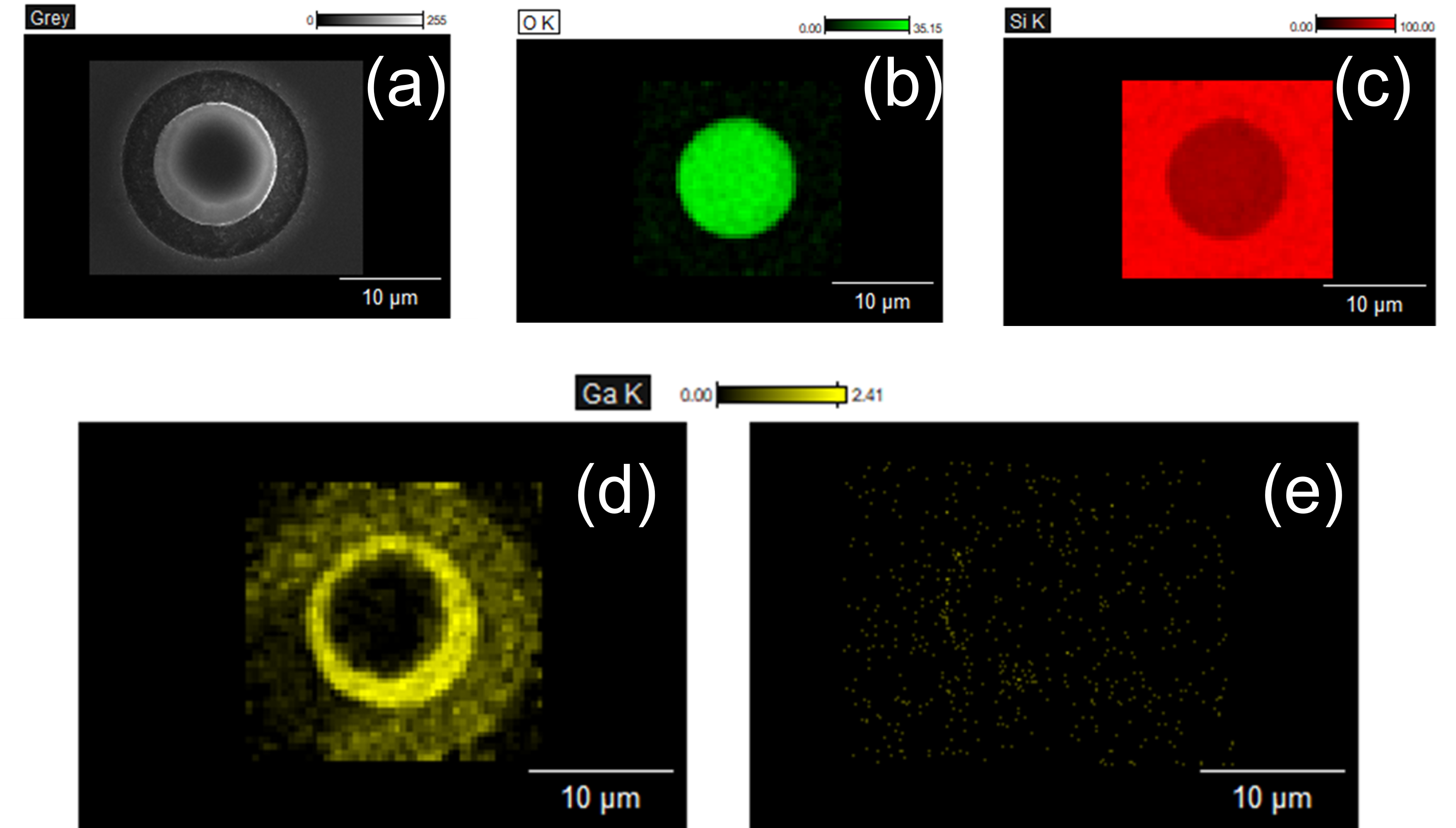}
    \caption{EDS plots of the microdisk with the (a) SEM image of the sample under analysis; elemental composition of (b) Oxygen  (c) Silicon (d) Ga ion before annealing (e) Ga ion post annealing }
    \label{fig:5}
\end{figure*}

The samples were then polished using a commercial 30 kV dual-beam FIB system (Thermo Fisher Helios 5 UC). For this Thermo Fisher system, the electron gun is normal and the ion gun is oriented at 38$^\circ$ to the stage, and the sample is tilted by 52$^\circ$ towards the ion gun during patterning. In this system, a 30 kV Gallium liquid metal ion source FIB column is combined with a Thermal Schottky Field-emitter scanning electron microscope gun (oriented at 90° to the sample) in an ultra-high vacuum chamber. The ability to image the sample with SEM ensures that the rough sidewall is milled deterministically with high precision. The pattern of pulses was controlled programmatically. The total ion dose at each feature was controlled through the beam exposure time and the beam current. The optimal process parameters for polishing are listed in Table \ref{tab:table_2}. We limit the beam current to lower values to mitigate the effects of Ga ion implantation into the SiO\textsubscript{2} surface. While higher beam currents could facilitate faster removal of roughness or material, they also lead to deeper Ga ion implantation and potential surface damage. In contrast, lower beam currents offer more precise material removal, ensuring finer control over the polishing process with minimal unintended modifications to the SiO\textsubscript{2} layer. The microdisk with sidewalls polished using the optimised FIB parameters is shown in Figure \ref{fig:4}.

\begin{table}[H]
    \centering
    \begin{adjustbox}{max width=\textwidth} % Adjusts table to fit within text width
    \begin{tabular}{|c|c|c|}
        \hline
        \textbf{Parameter} & \textbf{Value} \\ \hline
        Ion Source (Beam Type) & Ga\textsuperscript{+} \\ \hline
        Beam Current & 2.5 nA \\ \hline
        Accelerating Voltage & 30 kV \\ \hline
        Dwell time & 1 $\mu$s \\ \hline
        Sample Tilt Angle & 52° \\ \hline 
    \end{tabular}
    \end{adjustbox}
    \caption{Process Parameters for Focused Ion Beam polishing}
    \label{tab:table_2}
\end{table}

\subsection{\label{sec:level2}Effect of Annealing}

Ga is said to have a very limited solubility in silicon dioxide \cite{Diffusion_data} as the diffusion rates in silicon dioxide tend to be slower, as SiO\textsubscript{2} is more resistant to ion transport due to its amorphous and tightly packed structure. Gallium ions, like many other metals, experience low diffusion in SiO\textsubscript{2} at standard conditions. Additionally, the diffusion profile is influenced by whether the SiO\textsubscript{2} film was deposited through chemical vapor deposition (CVD) or grown thermally, as these methods impact the density of SiO\textsubscript{2} and defect concentration, thereby affecting ion migration. To release trapped gallium (Ga) ions from silicon dioxide (SiO\textsubscript{2}), annealing conditions can have a significant impact. Annealing in a vacuum or an inert atmosphere like nitrogen is typically preferred for out-diffusion of metals, including Ga, as it minimizes oxidation and may allow better mobility for the trapped ions to migrate to the surface and desorb.

Annealing in an oxygen-rich environment, on the other hand, could lead to further oxidation of Ga ions or create new trapping sites within the SiO\textsubscript{2} matrix due to the formation of Ga\textsubscript{2}O\textsubscript{3} or related oxides. This can make it more challenging to remove Ga, as these compounds might bind strongly within the SiO\textsubscript{2} structure. Vacuum annealing helps in promoting desorption of Ga by reducing the formation of these oxides, thus enhancing the likelihood of releasing Ga ions effectively \cite{Diffusion_data}.

\begin{figure*}[ht]
    \centering
    \includegraphics[width=1\linewidth]{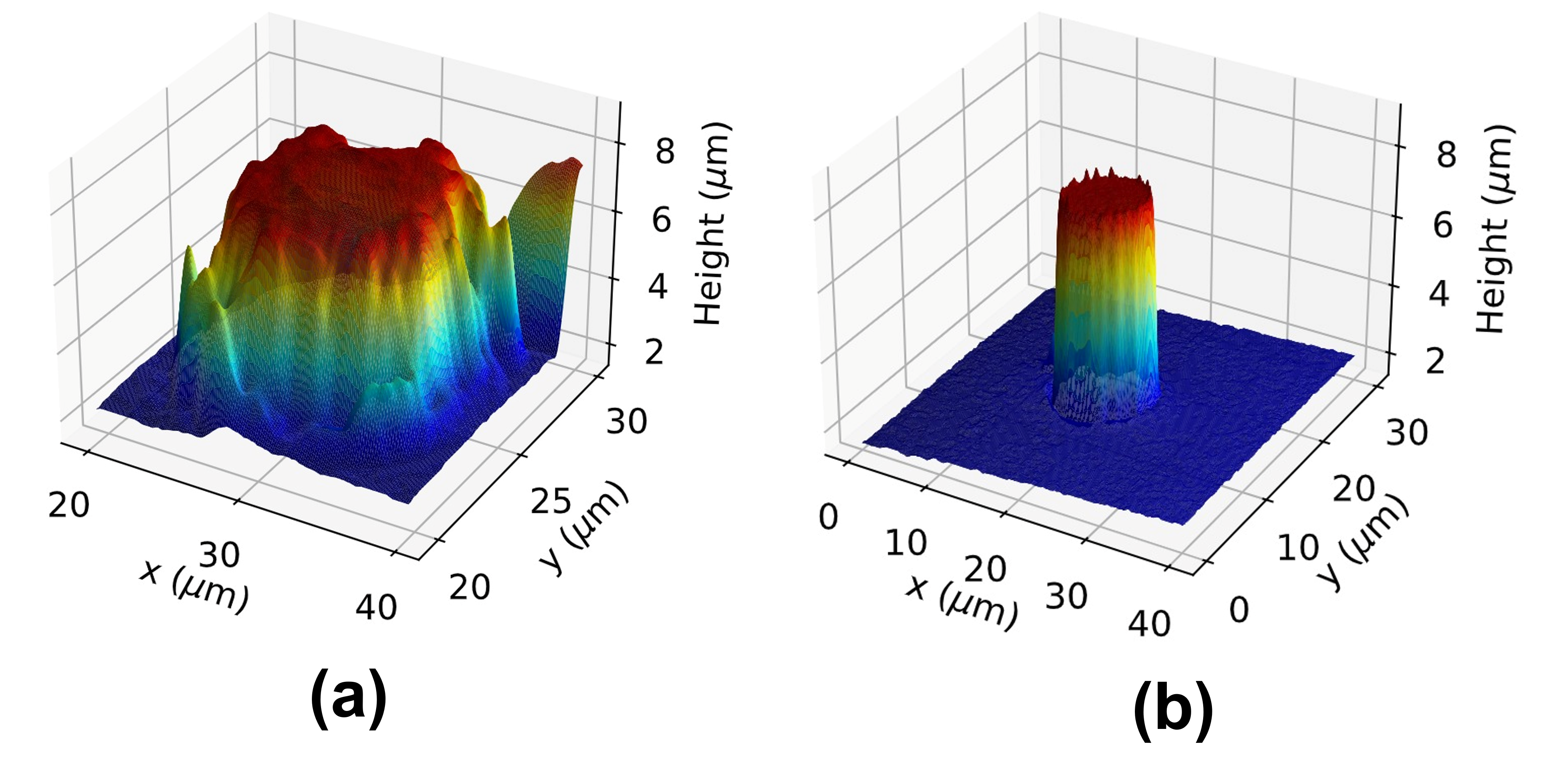}
    \caption{Optical 3D surface profiles of the (a) unpolished and (b) FIB polished microdisks}
    \label{fig:6}
\end{figure*}

Several studies and practices in semiconductor processing suggest vacuum annealing at elevated temperatures as a method to encourage out-diffusion and removal of metallic impurities from SiO\textsubscript{2} layers without risking oxidation that can further trap ions within the matrix. We anneal the polished microdisk sample in a vacuum environment using a furnace at 250°C. An elemental analysis of the sample through Energy Dispersive X-ray Spectroscopy (EDS)  (Figure \ref{fig:5} ) shows that the composition of Ga ions has drastically reduced in the polished sample after annealing. Thus, the annealing treatment effectively facilitated the release of trapped Ga ions from the microdisk resonator.

\section{Results and discussion}

The total quality factor $Q$ of a microdisk resonator comprises an intrinsic and an extrinsic term as given in Equation \ref{Eq:eq_1}. While the extrinsic term depends on the coupling mechanism, the intrinsic term is determined by the various intra-cavity loss mechanisms \cite{10.1063/1.117548}.  Equation \ref{Eq:eq_2} breaks down the intrinsic quality factor $Q_{\mathrm{in}}$ into into its individual loss mechanisms, including volume loss $Q_{\mathrm{vol}}$, surface loss $Q_{\mathrm{surf}}$ and radiative loss $Q_{\mathrm{rad}}$. These terms account for the various types of energy dissipation within the resonator material, such as scattering, absorption, and radiation, which collectively affect the intrinsic quality factor of the resonator. 
\begin{equation}
Q_{\mathrm{tot}}^{-1}=Q_{\mathrm{in}}^{-1}+Q_{\mathrm{ext}}^{-1}
\label{Eq:eq_1}
\end{equation}
\begin{equation}
Q_{\mathrm{in}}^{-1}=Q_{\mathrm{vol}}^{-1}+Q_{\mathrm{surf}}^{-1}+Q_{\mathrm{rad}}^{-1}
\label{Eq:eq_2}
\end{equation}

Surface roughness significantly influences the scattering quality factor of microresonators \cite{10.1063/1.2977681}. Numerical analysis on WGM microdisk resonators using FDTD simulations show that surface roughness along the microdisk perimeter significantly degrades the Q-factor, requiring it to be sub-100 nm for optimal performance \cite{Cho:11}. When light interacts with irregularities or variations on the surface of the microdisk, it can scatter rather than propagate smoothly. This scattering leads to energy loss from the resonator and reduces its ability to maintain a high Q factor. The quality factor can be framed in relation to surface roughness by considering how the roughness contributes to scattering losses. A common approach is to express the $Q$ factor as:
\begin{equation}
Q = \frac{\omega}{\Delta \omega} 
\label{Eq:eq_3}
\end{equation}
where \( \omega \) is the angular frequency of the resonant mode and \( \Delta \omega\) is the linewidth, which represents energy loss.

Surface roughness affects this relationship because increased surface irregularities lead to greater scattering losses, thereby increasing \( \Delta \omega\). This results in a lower $Q$ value. 

In more quantitative terms, models often relate surface roughness parameters (like root mean square height or correlation length) directly with scattering cross-sections that contribute to these losses. The specific dependence can vary based on factors like wavelength and material properties but generally indicates that smoother surfaces yield higher quality factors due to reduced scattering effects from imperfections.

\begin{figure*}[ht]
    \centering
    \includegraphics[width=1\linewidth]{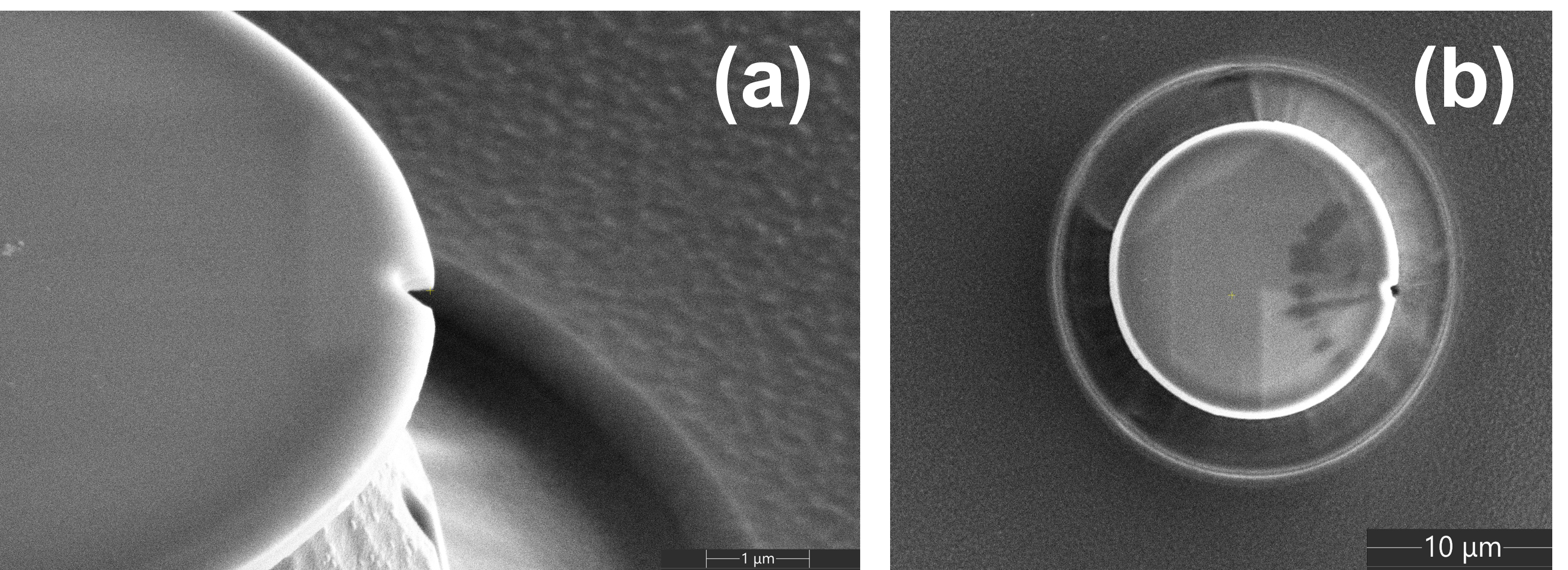}
    \caption{(a) Side-view and (b) top-view SEM images of notch integrated onto the polished microdisk using FIB}
    \label{fig:7}
\end{figure*}

A mathematical framework which incorporates the dielectric constant inhomogeneity and models surface roughness using variance and correlation length, leading to an expression for the quality factor (Q\textsubscript{ss}) that quantifies the impact of surface scattering losses as given in Equation \ref{Eq:eq_4} (\cite{Rahachou}, \cite{Borselli:05}):
\begin{equation}
Q_{\mathrm{SS}}=\frac{3 \lambda^3 D}{16 n \pi^2 \sigma^2 B^2}
\label{Eq:eq_4}
\end{equation}
where parameters including wavelength $\lambda$, diameter of microdisk $D$, surface roughness $\sigma$ and correlation length $B$ are shown to affect the scattering quality factor. The SiO\textsubscript{2} microdisk resonators are characterized using an optical profilometer to determine the effect of roughness on the quality of the resonators. White light interferometry (WLI) can rapidly measure large areas and obtain overall surface morphology, while Atomic Force Microscopy (AFM) is limited in its maximum evaluation length and is time-consuming for large-scale measurements \cite{YANG2020106200}. The white light interferometry mode allows quick positioning of areas of interest on large-scale surfaces, significantly reducing measurement time compared to AFM. For structures with steep gradients or large height differences, WLI can measure a wider range of surface profiles compared to AFM. While AFM may struggle with very steep slopes, WLI can capture these features more effectively \cite{LINDSETH1999276}. Additionally, WLI is non-contact, avoiding potential surface damage that could occur with AFM probes on delicate samples \cite{FANG2016297}. These features of WLI make it the appropriate choice for obtaining the surface profile of our microdisk structures which are isolated from the Si substrate with a significantly high pedestal. The surface profiles of the un-polished and polished microdisks of diameter $D$ of $\SI{5}{\um}$ are obtained using an optical 3-dimensional profilometer (Zeta 20) which works on the principle of white light interferometry. The root mean square values of surface roughness deduced for unpolished and polished microdisks are $ \SI{7}{nm}$ and $\SI{20}{nm}$, respectively. This translates to their corresponding Q\textsubscript{ss} being $3\times 10^2$ and $2\times 10^6$ respectively. Given that white light is used, an average wavelength of 650 nm is taken for calculations within the visible range. Thus, a giant enhancement of Q\textsubscript{ss} is achieved upon polishing the sidewalls of the microdisk using FIB. This method can also be extended for the effective integration of nano-holes or notches into resonators. Such perturbed systems offer several advantages, including efficient coupling of light into and out of the resonators \cite{PhysRevResearch.3.023202}, \cite{doi:10.1021/acsmaterialslett.4c00105} increased directionality of emission \cite{doi:10.1021/nl5048303}, and a lower lasing threshold when integrated with active materials \cite{Boost_lasing}. The integration of these local perturbations into the structure is challenging during patterning, as the addition of surface roughness in subsequent fabrication steps diminishes their intended effects. While polishing the microdisk completely eliminates roughness from the sidewalls, it effectively facilitates the fabrication of local perturbations, as shown in the Figure \ref{fig:7}.

\section{Conclusion}
In this work, we successfully fabricated and optimized SiO\textsubscript{2} microdisk resonators using conventional lithography techniques combined with FIB polishing. While the conventional fabrication process flow revealed significant sidewall roughness, particularly for smaller radius microdisk, FIB polishing was effectively implemented with optimized parameters for precise elimination of roughness. The FIB system may not be efficient for large-scale fabrication due to its inherently slow processing time (as the beam scans each individual element); however, this limitation is compensated by its exceptional precision. The study also mitigated Gallium ion implantation during FIB polishing through vacuum annealing at 250$^\circ$ C, as confirmed by Energy Dispersive X-ray Spectroscopy (EDS) analysis. A comprehensive analysis of factors affecting the quality factor of microdisk resonators was discussed, focusing on the impact of surface roughness on scattering losses. This work advances microdisk resonator fabrication techniques, offering a promising approach to enhance device performance for various applications in photonics. The combination of conventional methods with FIB polishing and subsequent annealing presents a novel strategy for improving the quality and functionality of microdisk resonators.\\

\emph{Acknowledgement.} A.K. acknowledges funding from the Department of Science and Technology via the grants: SB/S2/RJN-110/2017, ECR/2018/001485 and DST/NM/NS-2018/49.

\bibliography{references.bib}% Produces the bibliography via BibTeX.

\end{document}